\documentclass[letterpaper,oneside,11pt]{article} 

\usepackage{amssymb}
\usepackage{amsmath}
\usepackage{amsthm}
\usepackage{amstext}

\theoremstyle{plain}
\newtheorem{theorem}{Theorem}[section]
\newtheorem{lemma}[theorem]{Lemma}

\theoremstyle{definition}

\newcommand*{\dd}{{\mathrm{d}}}

\newcommand*{\bN}{{\mathbb{N}}}
\newcommand*{\bR}{{\mathbb{R}}}

\newcommand*{\cD}{{\mathcal{D}}}
\newcommand*{\cZ}{{\mathcal{Z}}}

\newcommand*{\pb}{{\bar{p}}}
\newcommand*{\rb}{{\bar{r}}}

\newcommand*{\st}{{\tilde{s}}}
\newcommand*{\zone}{z_1}
\newcommand*{\ztwo}{z_2}
\newcommand*{\zt}{z}
\newcommand*{\alphat}{{\tilde{\alpha}}}
\newcommand*{\betat}{{\tilde{\beta}}}

\newcommand*{\dergn}[2]{{{ \textstyle \frac{\mathrm{d}}{\mathrm{d}{#2}}}{#1}}}
\newcommand*{\derg}[2]{{{ \textstyle \frac{\mathrm{d}}{\mathrm{d}{#2}^{\downarrow}}}{#1} }}
\newcommand*{\der}[3]{{ \left. \derg{#1}{#2} \right|_{{#2}={#3}} }}
\newcommand*{\dist}{\delta}

\title{On the Variational Distance of Independently Repeated Experiments}

\author{Renato Renner \\
  ETH Z\"urich; Switzerland \\ {\tt renner@inf.ethz.ch}}

\date{}

\begin{document}

\maketitle

\begin{abstract}
  Let $P$ and $Q$ be two probability distributions which differ only
  for values with probability at least $p>0$. We show that the
  variational distance $\dist(P^n,Q^n)$ between the $n$-fold product
  distributions $P^n$ and $Q^n$ is upper bounded by~$\sqrt{n/(2 p)}\,
  \dist(P,Q)$, i.e., it cannot grow faster than the square root of
  $n$.
\end{abstract}

\section{Preliminaries}

Let $P$ be a probability distribution with range $\cZ$ and let $n \in
\bN$. We denote by $P^n$ the $n$-fold \emph{product
  distribution}, that is,
\[
  P^n(z_1, \ldots, z_n) = \prod_{i=1}^n P(z_i) 
\]
for any $z_1, \ldots, z_n \in \cZ$. Note that $P^n$ describes $n$
independently repeated random experiments with distribution $P$.

The \emph{variational distance} between two probability distributions
$P$ and $Q$ with range $\cZ$ is defined as\footnote{See, e.g.,
  \cite{CovTho91}. $\dist(\cdot, \cdot)$ is also called
  \emph{statistical difference}~\cite{Reyzin04}, \emph{Kolmogorov
    distance}, or \emph{trace distance}~\cite{NieChu00}.}
\[
\dist(P,Q) := \frac{1}{2} \sum_{z \in \cZ} |P(z) - Q(z)| \ .
\]
Note that $\dist$ is a distance measure on the set of probability
distributions with range $\cZ$. In particular, $\dist$ is symmetric,
$\dist(P,Q) = 0$ if and only if $P=Q$, and the triangle inequality
\begin{equation} \label{eq:tri}
  \dist(P, Q) \leq \dist(P, P') + \dist(P', Q)
\end{equation}
holds.

\section{Main Result and Proof}

\subsection{Upper Bounds for the Variational Distance}

Let $P$ and $Q$ be two probability distributions with range $\cZ$.  It
is a direct consequence of the triangle inequality~\eqref{eq:tri} that
the variational distance $\dist(P^n, Q^n)$ between the $n$-fold
product distributions $P^n$ and $Q^n$ cannot grow faster than linearly
in $n$, i.e.,
\begin{equation} \label{eq:linear}
  \dist(P^n, Q^n) \leq n \dist(P, Q) \ .
\end{equation}
Moreover, it is easy to find examples where this inequality is almost
tight. Let, e.g., $P$ and $Q$ be two binary distributions with range
$\cZ = \{0,1\}$ such that $P(1)=0$ and $Q(1) = \varepsilon$ for some
$\varepsilon>0$. If $n \varepsilon \ll 1$ then the variational
distance $\dist(P^n, Q^n)$ is roughly equal to $n \dist(P, Q) = n
\varepsilon$.

However, the upper bound~\eqref{eq:linear} can only be close to
optimal if, for some element $z \in \cZ$, the relative difference
$|P(z)-Q(z)|/(P(z) + Q(z))$ between the probabilities is large. (Note
that, in the above example, this relative difference for $z=1$ equals
one.)  Indeed, the following result states that, in all other cases,
$\dist(P^n, Q^n)$ cannot grow more than the square root of~$n$.

\begin{lemma} \label{lem:int}
  Let $P$ and $Q$ be two probability distributions with range $\cZ$,
  let $\cD := \{z \in \cZ : P(z) \neq Q(z)\}$ be the subset of $\cZ$
  where $P$ and $Q$ differ, and let $\pb:=\inf_{z \in \cD}(\min(P(z),
  Q(z)))$. If $\pb>0$ then, for any $n \in \bN$,
  \[
    \dist(P^n, Q^n) 
  \leq 
      \sqrt{\frac{1}{\pi \pb}}  \sqrt{n+\frac{1}{\pb}} 
    \, \dist(P, Q)
  \]
  and
   \[
    \dist(P^n, Q^n) 
  \leq 
      \sqrt{\frac{n}{2 \pb}} \, \dist(P, Q) \ .
  \] 
\end{lemma}

The first bound of Lemma~\ref{lem:int} is optimal in the sense that,
for any $\pb>0$, there are probability distributions $P$ and $Q$ with
minimum probability $\pb$ such that the quotient between the left and
the right hand side of the inequality approaches one for increasing
$n$ (as long as $\dist(P^n,Q^n) \ll 1$). On the other hand, the
constant $\frac{1}{2}$ in the second bound is the smallest value of
$c$ such that $\dist(P^n, Q^n) \leq \sqrt{c n/{\pb}} \, \dist(P, Q)$
always holds.

The result of Lemma~\ref{lem:int} is also complementary to a lower
bound derived in~\cite{Reyzin04}, which states that, if the maximum
probabilities of $P$ and $Q$ are small enough and if $n$ is not too
large, then $\dist(P^n, Q^n) \geq \Omega(\sqrt{n}) \dist(P,Q)$.

\subsection{Proof of Lemma~\ref{lem:int}}

To prove Lemma~\ref{lem:int}, we consider a probability distribution
$P_t$ parameterized by some real value $t$ and compute a bound on the
derivative, with respect to $t$, of the variational distance
$\dist(P_t^n, P_{t_o}^n)$.


\begin{lemma} \label{lem:der}
  Let $\{P_{t}\}_{t \in \bR}$ be a family of probability distributions
  with range $\cZ$ parameterized by $t \in \bR$, let $t_0 \in \bR$,
  and let $\zone, \ztwo \in \cZ$ such that\footnote{$\der{
      f(t)}{t}{t_0}$ denotes the right derivative $\lim_{t \to
      t_0^{+}} \frac{f(t) - f(t_0)}{t-t_0}$ of the function $f$ at
    $t=t_0$.}
  \[
    \der{P_t(\zt)}{t}{t_0}
  = 
    \begin{cases}
      1 & \text{if $\zt = \zone$} \\
      -1 & \text{if $\zt = \ztwo$}
    \end{cases} 
  \]
  and $P_t(\zt) = P_{t_0}(\zt)$ for any $\zt \in \cZ \backslash
  \{\zone, \ztwo\}$.  If $p:=P_{t_0}(\zone)$ and $p':=P_{t_0}(\ztwo)$
  are nonzero then
  \[
     \der{\dist(P_{t}^n, P_{t_0}^n)}{t}{t_0} 
  \leq
      \frac{1}{\sqrt{2 \pi}} 
    \sqrt{\frac{1}{p} + \frac{1}{p'}} 
    \, \sqrt{n+\frac{1}{\min(p,p')}}
  \]
  and
  \[
     \der{\dist(P_{t}^n, P_{t_0}^n)}{t}{t_0} 
  \leq
      \frac{1}{2} 
    \sqrt{\frac{1}{p} + \frac{1}{p'}} 
    \, \sqrt{n} \ .
  \]
\end{lemma}

\begin{proof}
  Assume without loss of generality that $t_0 = 0$. Since $P_t(\zt)$
  does not depend on $t$ for $\zt \in \cZ \backslash \{\zone,
  \ztwo\}$, we have $P_t(\zone) + P_t(\ztwo) = p + p'$, and thus, by
  the definition of the variational distance,
  \[
    \dist(P_{t}^n, P_{0}^n)
  =
    \frac{1}{2} \sum_{k=0}^{n} 
      q(k) \sum_{r=0}^{k} \binom{k}{r} | a_{k,r}(t) |
  \]
  where
   \[
    q(k) := \binom{n}{k} (p+p')^k (1-p-p')^{n-k}
  \]
  and
  \[
    a_{k,r}(t)
  :=
      \left(\frac{p}{p+p'}\right)^r
        \left(\frac{p'}{p+p'}\right)^{k-r}
    - \left(\frac{P_{t}(\zone)}{p+p'}\right)^r
        \left(\frac{P_{t}(\ztwo)}{p + p'}\right)^{k-r} 
     \ .
  \]
  Let $a'_{k,r}(0):=\der{a_{k,r}(t)}{t}{0}$.  Because $a_{k,r}(0)=0$,
  we have
  \[
  \der{|a_{k,r}(t)|}{t}{0} = |a'_{k,r}(0)| \ .
  \]  
  Moreover, it follows from~\eqref{eq:derpot} that $a'_{k,r}(0) \geq
  0$ if and only if $r \leq \rb := \lfloor k \frac{p}{p+p'} \rfloor$.
  We thus conclude
  \begin{equation} \label{eq:completesum}
    \der{\dist(P_{t}^n, P_{0}^n)}{t}{0} 
  = 
    \sum_{k=0}^{n} q(k)
      \sum_{r=0}^{\rb} \binom{k}{r}
        a'_{k,r}(0) \ .
  \end{equation}
  Let $\alpha:=\frac{p}{p+p'}$ and $\beta:=\frac{p'}{p+p'}$. The last
  sum in the above expression can then be rewritten as
  \begin{align}
    \sum_{r=0}^{\rb} \binom{k}{r}
      a'_{k,r}(0)
  & = 
    \frac{1}{p+p'} \sum_{r=0}^{\rb} 
    \left[
      \binom{k}{r} (k-r) \alpha^r \beta^{k-r-1}
      - \binom{k}{r} r \alpha^{r-1} \beta^{k-r}
    \right] \nonumber \\
  & = 
    \frac{k}{p+p'} 
      \left[
        \sum_{r=0}^{\rb} 
          \binom{k-1}{r} \alpha^r \beta^{k-r-1}
        - \sum_{r=0}^{\rb-1} 
          \binom{k-1}{r} \alpha^r \beta^{k-r-1} 
      \right] \nonumber \\
  & =
    \frac{k}{p+p'} 
    \binom{k-1}{\rb} \alpha^\rb  \beta^{k-\rb-1} \ .  \label{eq:intsum}
  \end{align}
  With the definition $\alphat:=\frac{\rb+1}{k+1}$,
  $\betat:=\frac{(k+1)-(\rb+1)}{k+1}$, we have
  \begin{align}
    \binom{k-1}{\rb} \alpha^\rb \beta^{k-\rb-1}
  & =  
    \binom{k+1}{\rb+1} \frac{(\rb+1) ((k+1)-(\rb+1))}{k (k+1)}
      \frac{\alpha^{\rb+1} \beta^{(k+1)-(\rb+1)}}{\alpha \beta} \nonumber \\
  & \leq
    \binom{k+1}{\rb+1} \frac{(\rb+1) ((k+1)-(\rb+1))}{k (k+1)} 
      \frac{\alphat^{\rb+1} \betat^{(k+1)-(\rb+1)}}{\alpha \beta} \nonumber\\
  & \leq
    \sqrt{\frac{1}{2 \pi (k+1) \alphat \betat}}
    \cdot \frac{(\rb+1) ((k+1)-(\rb+1))}{k (k+1)\alpha \beta} \nonumber\\
  & = 
    \frac{1}{k} \sqrt{\frac{k+1}{2 \pi \alpha \beta}} \,
      \sqrt{\frac{\alphat \betat}{\alpha \beta}} \ , \label{eq:intbinombound} 
  \end{align}
  where the first inequality follows from~\eqref{eq:maxpot} and the
  second from Lemma~\ref{lem:stirbin}.  Using the definition of $\rb$
  and letting $\gamma:=k \alpha - \lfloor k \alpha \rfloor$, the
  expression in the second square root of the last term can be bounded
  by
  \begin{multline}
    \frac{\alphat \betat}{\alpha \beta}
  =    
    \frac{\lfloor k \alpha \rfloor+1}{\alpha (k+1)} 
      \cdot \frac{(k+1) - (\lfloor k \alpha \rfloor +1)}{ \beta (k+1)}
  =
    \frac{k \alpha+(1-\gamma)}{\alpha (k+1)} 
      \cdot \frac{k \beta + \gamma}{\beta (k+1)} \\
  =
    \frac{k + \frac{1-\gamma}{\alpha}}{k+1} 
    \cdot \frac{k + \frac{\gamma}{\beta}}{k+1} 
  \leq
    \frac{k+\frac{1}{\min(\alpha, \beta)}}{k+1} \ ,
  \end{multline}
  where the last inequality follows from the fact that
  $\frac{1-\gamma}{\alpha} = \frac{1-\gamma}{1-\beta}$ and
  $\frac{\gamma}{\beta}$ cannot both be larger than one, since $\beta,
  \gamma \in [0,1]$.  Combining this with~\eqref{eq:intbinombound}
  and~\eqref{eq:intsum}, we find
  \begin{equation} \label{eq:partsumone}
    \sum_{r=0}^{\rb} \binom{k}{r}
      a'_{k,r}(0)  
  =
    \frac{k}{p+p'} \binom{k-1}{\rb} \alpha^\rb \beta^{k-\rb-1}
  \leq 
    s(k)
    \ ,
  \end{equation}
  where
  \[
    s(k)
  :=   
      \frac{1}{p+p'} 
    \sqrt{\frac{k+\frac{1}{\min(\alpha, \beta)}}{2 \pi\alpha \beta}} \ .
  \]
  Alternatively, the left hand side of~\eqref{eq:partsumone} can be
  upper bounded by
  \begin{equation} \label{eq:salt}
    \st(k)
  :=   
    \frac{1}{p+p'} \, \sqrt{\frac{k}{4 \alpha \beta}} \ .
  \end{equation} 
  To see this, assume first that $\alpha k \geq 2$ and $\beta k \geq
  2$. Then $\frac{1}{\min(\alpha, \beta)} \leq \frac{k}{2}$ which
  implies
  \[
    s(k) 
  \leq 
    \frac{1}{p+p'} \sqrt{\frac{\frac{3}{2} k}{2 \pi \alpha \beta}} 
  <
    \st(k) \ .
  \]
  On the other hand, if $\alpha k < 2$ or $\beta k < 2$, the bound
  follows from a straightforward calculation using~\eqref{eq:maxpot}.
  (In this case, $\rb$ or $(k-1)-\rb$ is either $0$ or $1$, i.e., the
  binomial $\binom{k-1}{\rb}$ in~\eqref{eq:partsumone} equals $1$ or
  $k-1$.)
  
  When~\eqref{eq:completesum} is combined with~\eqref{eq:partsumone},
  we obtain
  \begin{equation} \label{eq:sumcompletebound}
    \der{\dist(P_t^n, P_{0}^n)}{t}{0} 
  \leq
    \sum_{k=0}^{n} q(k) s(k) \ .
  \end{equation}
  Note that $s(k)$ is a concave function in $k$ and that $q(k)$ are
  the probabilities of a binomial distribution with mean $n (p+p')$,
  that is, $\sum_{k=0}^n q(k) = 1$ and $\sum_{k=0}^n q(k) k = n
  (p+p')$.  We can thus apply Jensen's inequality to find an upper
  bound for the sum on the right hand side
  of~\eqref{eq:sumcompletebound}, i.e.,
  \[
    \der{\dist(P_t^n, P_0^n)}{t}{0} 
  \leq
    \sum_{k=0}^{n} q(k)  s(k)
  \leq
    s\bigl(\sum_{k=0}^{n} q(k) k\bigr) 
  = 
    s(n (p+p')) \ ,
  \]
  from which the first inequality of the lemma follows.
  
  Similarly, because $\st(k)$ is concave in $k$ as well, we have
  \[
    \der{\dist(P_t^n, P_0^n)}{t}{0} 
  \leq
    \st(n  (p+p')) \ ,
  \]
  which implies the second inequality of the lemma.
\end{proof}

We now use the bounds provided by Lemma~\ref{lem:der} to prove our
main result.

\begin{proof}[Proof of Lemma~\ref{lem:int}]
  
  We first prove the first inequality of the lemma for the special
  case where the probabilities $P$ and $Q$ only differ for two values
  $\zone$ and $\ztwo$. Assume (without loss of generality) that
  $P(\zone) \leq Q(\zone)$ and let $p:=P(\zone)$, $p':= P(\ztwo)$,
  $q:=Q(\zone)$.  For $t \in [p,q]$, let $P_t$ be the distribution
  with range $\cZ$ given by
\[
  P_t := \frac{q-t}{q-p} P + \frac{t-p}{q-p} Q \ .
\]
i.e., $P_t(\zone) = t$, $P_t(\ztwo) = p+p'-t$, and $P_t(\zt) = P(\zt)
= Q(\zt)$ for any $\zt \in \cZ \backslash \{\zone, \ztwo\}$. We can
thus apply the first inequality of Lemma~\ref{lem:der} which gives
\[
\begin{split}
  \der{\dist(P_s^n, P_{t}^n)}{s}{t} 
& \leq       
       \frac{1}{\sqrt{2 \pi}} 
    \sqrt{\frac{1}{t} + \frac{1}{p+p'-t}} 
    \, \sqrt{n+\frac{1}{\min(t,p+p'-t)}} \\
& \leq
      \frac{1}{\sqrt{2 \pi}} 
    \sqrt{\frac{2}{\pb}} 
    \, \sqrt{n+\frac{1}{\pb}}   \ .
\end{split}
\]
Using Lemma~\ref{lem:pathint}, we obtain
\[
  \dist(P^n, Q^n) = \dist(P_p^n, P_q^n) 
\leq
  \int_{p}^{q}  \der{\dist(P_s^n, P_{t}^n)}{s}{t} \dd t
\leq
  (q-p) \frac{1}{\sqrt{2 \pi}} 
    \sqrt{\frac{2}{\pb}} 
    \, \sqrt{n+\frac{1}{\pb}} \ .
\]
Since $q-p = \dist(P,Q)$, this concludes the proof of the first
inequality of the lemma for the special case where the probabilities
$P$ and $Q$ differ for at most two values in $\cZ$.

To prove the general case, we first observe that if the set $\cD$ is
infinite, then $\pb$ equals zero and nothing has to be proven. On the
other hand, if $\cD$ is finite, it is easy to see that there exists a
sequence $(P_i)_{i=1, \ldots, m}$ (for some $m \in \bN$) of
distributions with range $\cZ$ such that
\begin{itemize}
\item $P_1 = P$ and $P_m = Q$,
\item for any $i \in \{1, \ldots, m\}$, the distributions $P_i$ and
  $P_{i+1}$ differ only for two elements in $\cD$,
\item $\min_{z \in \cD} P_i(z) \geq \pb$, for all $i \in \{1, \ldots,
  m\}$, and
\item $\sum_{i=1}^{m-1} \dist(P_i, P_{i+1}) = \dist(P_1, P_m)$.
\end{itemize}
The general assertion then follows directly from the special case
proven above and the triangle inequality~\eqref{eq:tri}, i.e.,
\[
\begin{split}
  \dist(P^n, Q^n)
\leq 
  \sum_{i=1}^{m-1} \dist(P_i^n P_{i+1}^n)
& \leq
  \sum_{i=1}^{m-1} \sqrt{\frac{1}{\pi \pb}} \, \sqrt{n+\frac{1}{\pb}} 
    \, \dist(P_i, P_{i+1}) \\
& =
  \sqrt{\frac{1}{\pi \pb}} \, \sqrt{n+\frac{1}{\pb}} 
    \, \dist(P, Q) \ .
\end{split}
\]

The second inequality follows by exactly the same reasoning based on
the second inequality of Lemma~\ref{lem:der}.
\end{proof}

\appendix

\section{Appendix: Some Useful Identities}

\subsection{An Upper Bound for the Binomial Coefficient}

\begin{lemma} \label{lem:stirbin}
  For any $n \geq k > 0$,
\[
    \binom{n}{k} 
  \left(\frac{k}{n}\right)^k 
  \left(\frac{n-k}{n}\right)^{n-k} 
\leq
  \sqrt{\frac{n}{2 \pi k (n-k)}} \ .
\]
\end{lemma}

\begin{proof}
  The assertion follows directly from Stirling's
  approximation~\cite{Feller68Stirling},
\[
  \sqrt{2 \pi} n^{n+1/2} e^{-n+1/(12n+1)} 
< 
  n! 
<
  \sqrt{2 \pi} n^{n+1/2} e^{-n+1/(12n)} \ 
\]
for $n>0$.
\end{proof}

\subsection{Bounding the Variational Distance by Its Path Integral}

\begin{lemma} \label{lem:pathint}
  Let $a<b$, let $\{P_t\}_{t \in \bR}$ be a family of probability
  distributions parameterized by $t \in [a,b]$, and let
  $f(t):=\der{\dist(P_{s}, P_{t})}{s}{t}$ be the right derivative of
  the variational distance. Then
\[
  \dist(P_b, P_a) \leq \int_{a}^{b} f(t) \dd t \ .
\]
\end{lemma}

\begin{proof}
Since equality holds for $b=a$, it suffices to verify that
\begin{equation} \label{eq:distder}
  \derg{\dist(P_{r}, P_a)}{r} \leq \derg{\int_{a}^{r} f(t) \dd t}{r} \ ,
\end{equation}
for any $r \in (a,b)$. Using the triangle inequality for the
variational distance, the expression on the left hand side can be
bounded by
\[
  \derg{\dist(P_{r}, P_a)}{r}
=
  \lim_{\varepsilon \to 0^+} 
    \frac{\dist(P_{r+\varepsilon}, P_a) -\dist(P_{r}, P_a)}{\varepsilon}
\leq
  \lim_{\varepsilon \to 0^+} 
    \frac{\dist(P_{r+\varepsilon}, P_{r})}{\varepsilon}
=
  f(r) \ .
\]
Inequality~\eqref{eq:distder} then follows from the second fundamental
theorem of calculus, which concludes the proof.
\end{proof}

\subsection{Auxiliary Identities}

For any $n>0$, $k \in [0, n]$, and $x \in [0,1]$,
\begin{equation} \label{eq:derpot}
  \dergn{x^k (1-x)^{n-k}}{x}
\geq 0 \quad \Longleftrightarrow \quad
  x \leq \frac{k}{n} \ .
\end{equation}
As an immediate consequence of this expression we have
\begin{equation} \label{eq:maxpot}
  x^k (1-x)^{n-k} 
\leq 
  \left(\frac{k}{n}\right)^k \left(1-\frac{k}{n}\right)^{n-k} \ .
\end{equation}

\end{document}